\documentclass[aps, pra,twocolumn,showpacs,amsmath,amssymb,superscriptaddress, longbibliography, 10pt]{revtex4-1}

\usepackage{tgtermes}
\usepackage[utf8]{inputenc}
\usepackage{bm}
\usepackage{color}
\usepackage{makeidx}
\usepackage{amsfonts}
\usepackage{amssymb}
\usepackage{amsthm}
\usepackage{graphicx}
\usepackage{epstopdf}
\usepackage{subfigure}
\usepackage{amsmath,amssymb}
\usepackage[colorlinks=true,citecolor=blue,urlcolor=blue]{hyperref}
\usepackage{color}

\newcommand{\bea}{\begin{eqnarray}}
\newcommand{\ea}{\end{eqnarray}}
\newcommand{\eea}{\end{eqnarray}}

\newcommand{\sumint}[1]

\begin{document}

\title{Quantum reflection of a Bose-Einstein condensate from a rapidly varying potential: the role of dark soliton }
\author{Dongmei Wang}
\affiliation{Institute of Modern Physics, Northwest University, Xi'an, 710127, China}
\author{Bo Xiong}
\email{boxiongpd@gmail.com}
\affiliation{School of Science, Wuhan University of Technology, Wuhan 430070, China}
\author{Tao Yang}
\email{yangt@nwu.edu.cn}
\affiliation{Institute of Modern Physics, Northwest University, Xi'an, 710127, China}
\affiliation{Shaanxi Key Laboratory for Theoretical Physics Frontiers, Xi'an, 710127, China}
\affiliation{NSFC-SPTP Peng Huanwu Center for Fundamental Theory, Xian 710127, China}

\date{\today}

\begin{abstract}
   We study the dynamic behavior of a Bose-Einstein condensate (BEC) containing a dark soliton separately reflected from potential drops and potential barriers. It is shown that for a rapidly varying potential and in a certain regime of incident velocity, the quantum reflection probability displays the cosine of the deflection angle between the incident soliton and the reflected soliton, i.e., $R(\theta)$ $\sim $ $\cos 2\theta$. For a potential drop, $R(\theta)$ is susceptible to the widths of potential drop up to the length of the dark soliton and the difference of the reflection rates between the orientation angle of the soliton $\theta=0$ and $\theta=\pi/2$, $\delta R_s$, displays oscillating exponential decay with increasing potential widths. However, for a barrier potential,  $R(\theta)$ is insensitive for the potential width less than the decay length of the matter wave and $\delta R_s$ presents an exponential trend. This discrepancy of the reflectances in two systems is arisen from the different behaviors of matter waves in the region of potential variation.
\end{abstract}


\maketitle
\section{Introduction}

Quantum reflection is a classically counterintuitive phenomenon whereby particles are reflected from potential barriers or wells without reaching a classical turning point \cite{Friedrich:2004a, Landau:1992a}.
Such phenomenon has been observed in various systems, such as atomic mirror \cite{PhysRevLett.77.1464,Segev:1997a,C1998Retardation}, nanoporous media \cite{PhysRevLett.97.093201,Dufour:2013a}, rough surfaces \cite{Zhao:2010a}, and Si, BK7 glass surfaces \cite{PhysRevLett.86.987}.
Recent developments are extended to cold atomic system where quantum reflections of ultra-cold atoms appear on the potential of solid surfaces \cite{PhysRevLett.95.073201,PRL.93.223201} as well as on the various structure, such as graphene \cite{Silvestre:2019a}, semiconductor heterostructures and thin films \cite{Judd:2011a}.
Since significent reflection occurs under the condition, $\phi(k)=(1/k^2)dk/dx_s \sim d\lambda/dx_s \gg 1$ \cite{PhysRevLett.86.987}, where the local wave number, $k=2\pi/\lambda$, depends on the de Broglie wave length, $\lambda$, and $x_s$ is the normal distance from the atom to the surface, the system of ultracold and quantum degenerate atomic gases with large de Broglie wave lengths is an excellent platform for experiments to study quantum reflection at normal incidence with large flexibility on the control of atomic motion.

Solitons are self-reinforcing wavepackets, which maintain its shape when propagating over long distances, and emerge from collisions unaltered \cite{Thie:2010a, Corn:2009a, Dodd:1982a}. The existence of such localized wavepackets indicates nontrivial effects from nonlinear interaction, and they have been observed in different systems, including water waves \cite{Hirota1973Exact}, plasmas \cite{Kuznetsov1987Soliton,Lonngren2000Soliton}, and nonlinear optics \cite{PhysRevLett.101.153904,Gibbon1973AnN}. Theoretically, solitons in nonlinear media were described numerically \cite{Zabusky1968Solitons} and analytically \cite{Newell1980The} half a century ago, and have been confirmed in different fields such as oceanography, biology, and fiber optics. Afterwards, the atomic solitons were also investigated in various platforms, such as optical systems \cite{Proukakis:2004a}, mangetic films \cite{Wu:2004a}, and waveguide arrays.
Recently , solitons in ultracold atomic system, e.g., Bose-Einstein condensate, have attracted much attention. Especially, such soliton can theoretically be interpreted well by the well-known Gross-Pitaevskii (GP) equations \cite{Burger:1999a, Denschlag:2000a}. Experimentally, a matter-wave soliton with a phase difference of $\pi$ can be generated readily by appropriate phase imprinting \cite{Denschlag:1999a,PhysRevLett.101.120406}, density engineering \cite{PhysRevLett.86.2926}, sweeping a dipole potential through the condensate \cite{PhysRevLett.99.160405,PhysRevLett.101.170404},
as a consequence of quantum shock, and from the local minima of interference fringes \cite{PhysRevLett.98.180401,PhysRevLett.101.130401}.

In recent years, a great deal of experimental and theoretical work has been involved with the quantum reflection of condensates. Conventionally, the incident velocity of the condensate has a significant effect on the reflection process, and when the velocity of the condensate is low, an abnormal reflectivity are observed experimentally \cite{PRL.93.223201,PhysRevLett.97.093201}. Moreover, apart from the light and single-atom reflections, the interaction between atoms is verified to be a key factor for atomic cloud reflection and diffraction \cite{PhysRevA.74.053605}. The nonlinear excitation in the condensate, such as solitons and vortices, arisen from interatomic interaction can destroy the cloud structure when the condensate interacts with the solid surface or the gaussian potential \cite{PhysRevA.74.043619,PhysRevA.75.065602}.
Due to the surprising robustness of bright matter-wave solitons, its collision and associated regions of chaotic dynamics have been addressed \cite{PhysRevLett.98.020402,PhysRevA.77.013620}. Also the quantum reflection of such bright solitons are discussed intensively \cite{2005Enhanced,Corn:2009a}. For matter-wave dark solitons, a recent theoretic work has shown that there exist some dissipation when they are reflected from soft walls \cite{PhysRevA.95.013628}. A latest work also shows that the orientation of a 2D dark soliton can affect significantly the reflection probability of the condensate from a sharp potential barrier\cite{Qiao:2018a}. However, the essential factors for such influence have not been presented. Moreover, the influence of a 2D dark soliton, especially its structure, subject to a potential well on the quantum reflection has not yet investigated. Correspondingly, the underlying mechanism of a BEC with a dark soliton reflected separately from potential well and potential barrier is still not clear. In this paper, we investigate the dynamic behavior of a 2D disk-shaped Bose-Einstein condensate containing a dark soliton separately reflected from the potential drop and the potential barrier. We try to find the influence of the orientation angle of the dark soliton, the structure of potential and the width of the potential on the quantum reflection rate. Afterwards, we further used dynamic images to reflect the physical mechanism through numerical simulation. The underlying mechanism of the quantum reflection is discussed intensively.

\section{method and Simulation scheme}
 We consider a BEC containing $N$ $^{23}$Na atoms, which are initially trapped by a harmonic trapping potential of the form $V_{h}(x, y, z) =
 m(\omega_{x}^{2} x^2 + \omega_{y}^{2} y^2 + \omega_{z}^{2} z^2)/2$. Here we chose
the trap frequencies to be $\omega = \omega_x = \omega_y = 2\pi \times 10{\rm Hz}$ and $\omega_z = 2\pi \times 100 {\rm Hz}$ so that the trap frequencies in the $z$ direction are much larger than that in the $x$ and $y$ directions. Hence, the system can be treated as a disk-shaped 2D condensate. The dynamics of such system at zero temperature is well described by the 2D time-dependent Gross-Pitaevskii equation,
\begin{equation}
   i\hbar \partial_t \psi = - \frac{\hbar^2}{2m} \psi + V(x, y)\psi + g_{2D} N|\psi|^{2}
\end{equation}
where $\psi(x,y,t)$ is the temporal condensate wave function and the number of atoms, $N$, is chosen to be $1.8\times10^4$ in our calculation. $g_{2D} = 2\sqrt{2\pi}\hbar \omega_z a_s a_z$ is the 2D coupling constant with the $s$ wave scattering length $a_s = 2.9$nm. The radial and axial oscillation lengths are $a_0 = \sqrt{\hbar/m\omega}$ and $a_z = \sqrt{\hbar/m\omega_z}$ respectively. The ground state of dark soliton is obtained by the imaginary time evolution of GP equation, where a $\pi$ phase step is compulsively added into the condensate wave function so that two stable density halves are formed. To get a prevalent insight about the interaction between dark soliton and potential surface, we consider general potentials  with the form,
\begin{equation}
   V(x,y) = \left\{ \begin{array}{cc} V_{h}(x, y, 0) &  x < 0\\
	                           h  &          x\in [0,d] \\
                               0    & x > d\end{array} \right.
\end{equation}
where $h>0$ and $h<0$ are the intensity of the potential step and potential drop, respectively. $d$ is the width of the barrier or well.

\begin{figure}[htbp]
\begin{center}
\includegraphics[angle=0,width=0.45\textwidth]{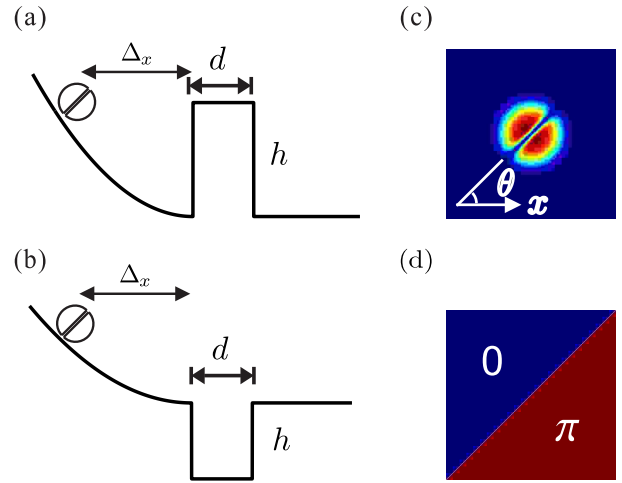}
\caption{A schematic view of a condensate containing a dark soliton, whose center-of-mass is away
                            from the potential step (a) and potential drop (b) with distance $\Delta_x$ in the $x$-axis. The top view of the initial density distribution of the condensate under one set of specific parameters ($\theta = \pi/4$) (c) and its correspond phase diagram (d).}
\label{fig:pz4}
\end{center}
\end{figure}

 We consider two different potential profiles: potential I [Fig.1 (a)] is the potential barrier of width $d$; potential II [Fig.1 (b)] is a potential well of width $d$. The sharp potential choosen here have fulfill readily the quantum reflection condition and make the reflection of solitons prominent. Thus it enables us to identify the key physical processes that occur when BEC experiences reflection, and thus sharpen the understanding needed for experimental studies to explain BECs with solitons reflected from the more complex semiconductor surface potential or nanostructure potential. In the middle of the harmonic trap, the center-of-mass of the condensate is placed at the position
which away from the trap center $(\Delta_x, 0)$. Due to the variation of trap potential, the condensate cloud is driven to move toward the region of lower potential where the potential barrier Fig.1(a) and potential drop Fig.1(b) occur, which allow us to study the reflection process of the condensate containing a dark soliton. Fig.1(c) and (d) show a typical density distribution and phase diagram of the initial state of a condensate with $\theta = \pi/4$
containing a dark soliton, respectively. When the solitons are initially imprinted in the condensate, we can freely
adjust the angle of orientation $\theta$ between the dark solitons and the positive $x$-axis. To derive the dimensionless GP equations, we use $x_0=\sqrt{\frac{\hbar}{m\omega}}$, $t_0 = 1/\omega$, $E_0 = \hbar\omega$ as unit of length, time, energy, respectively. Note that in the region of potential drop, the velocity of the condensate wave is much larger than the incident velocity so a sufficiently small grid size is desired to avoid the distortion arisen from finite size scale.

To explore the role of dark soliton in quantum reflection of BEC from the potentials, it is required that the lifetime of dark soliton should be larger than the timescale of reflection process. Since the dark soliton in 2D BEC is generally unstable, it will eventually decay into vortices due to snake instability. This indicates that the complete quantum reflection should appear before the distortion of the dark soliton. My latest work shows that the lifetime of dark soliton either located in the side area of 2D BEC or with quantum noise is still larger than the timescale of reflection process. Also we note that the stability of 2D dark soliton in trapped BEC is still an interesting topics on its own.

\section{Results and discussion}

As stated in \cite{PhysRevA.88.043602,PhysRevA.87.023603}, the center-of-mass velocity and interatomic interaction of condensates play an important role in the formation of interference and dynamic excitation. These two factors determine the time scales of interference and distortion as well as the competition of two time scales. Here, the quantum reflection of the condensate comes with quantum interference and possible nonlinear excitation, which indicates the center-of-mass velocity and interaction are also critical factors. As the condensate moves toward the trap center, its average incident velocity at the edge of the potential well or barrier is approximately, $\bar{v} \approx \omega\sqrt{\Delta_{x}^{2} + \Delta_{y}^{2}}$. Previous research shows that at low incident velocity and high density, the interference time of the BEC is extended while its distortion time is reduced significantly, as a result that dark solitons and votices are created in the process. Although the interaction has little effect on refection rate, it causes the fragmentation of the BEC and in the threshold of experimental measurement, the low reflection rate is obtained \cite{PhysRevLett.95.073201}.

Here, we further study the quantum reflection of BECs containing dark solitons from rapidly varying potentials, and specifically concentrate on the role of structure pattern of the nonlinear excitations in such reflection process. To demonstrate explicitly our findings, we define the reflection rate
as
\begin{equation}
R(t) = \frac{1}{N} \int_{-\frac{L}{2}}^{0} \int_{-\frac{L}{2}}^{\frac{L}{2}} |\psi(x,y,t)|^2 dx dy
\end{equation}
where $L$ is the system size chosen for calculations. We denote by $R_s$ the asymptotic reflection probability after the
condensate cloud has been completely reflected by the potential. When the height or depth of the potential, $|h|$, is enough
large, the BEC undergoes the elastic reflection with $R_s = 1$. In principle, the quantum reflection from two different potentials along $x$ direction is only dependent on the velocity in $x$ direction. While, for interacting system , the reflection definitely depend on the velocity in $y$ direction, i.e., $\omega_y\Delta_y$.

\begin{figure}[htbp]
\begin{center}
\includegraphics[angle=0,width=0.45\textwidth]{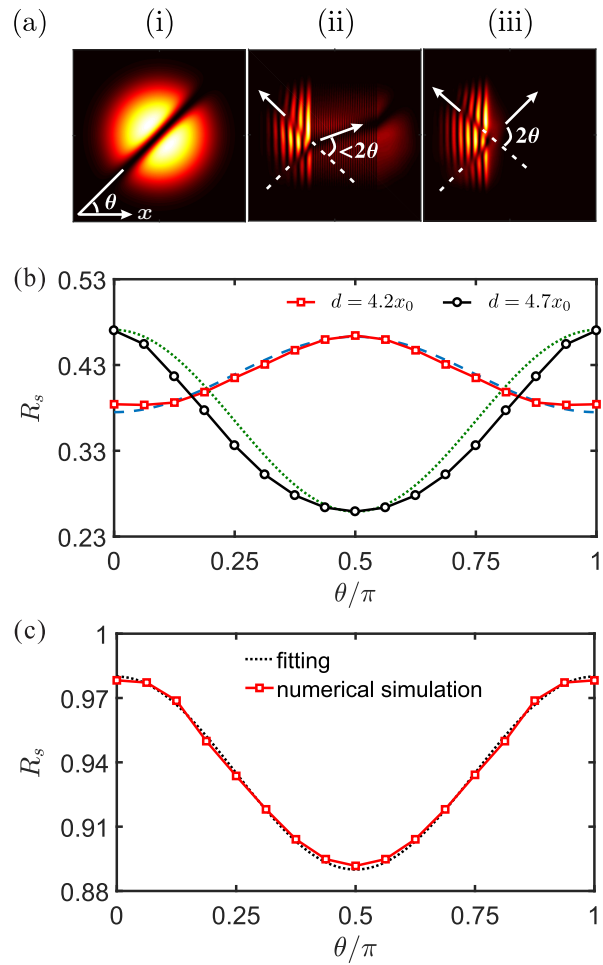}
\caption{(Color online) Dynamics of a BEC containing a dark soliton (a). The typical density plots show that the initial matter wave soliton with orientation angle $\theta$ (i) incidents separately on a potential drop (ii) and potential barrier (iii). The asymptotic reflection rate $R_s(\theta)$ of the BEC, reflected separately from potential drop (b) and potential step (c). The red curve with squares and the dark curve with circles in (b) are from numerical solutions and correspond to the width of the potential drop $d=4.2x_0$ and $d=4.7x_0$, respectively, and identical parameters: $h=-180E_0$, $\Delta_x=7.05x_0$. The line curves are the analytic fitting from the ansatz $R_s=R_0+R_{se}\cos2\theta$: $R_0=0.419$ and $R_{se}=-0.044$ (blue dashed line) and $R_0=0.365$ and $R_{se}=0.106$ (green dotted line). Similarly, the curve with squares in (c) is from numerical results for $d =5.64x_0$, $h = 40 E_0$, $\Delta_x = 7.05x_0$, and the solid line is from the ansatz for potential barrier with $R_0 = 0.935$ and $R_{se} = 0.045$}.
\label{fig:pz4}
\end{center}
\end{figure}

Our latest work shows that when a BEC with a dark soliton reflects from a potential barrier, its asymptotic reflection probability, $R_s$, is highly sensitive to the orientation angle $\theta$ of the soliton \cite{Qiao:2018a}. Here we further consider the behavior of such BEC, which is reflected from potential well. We find that although $R_s$ is still sensitive to the orientation angle in the system of potential well, the variation of $R_s$ as the function of $\theta$, i.e., $R(\theta)$, is strongly dependent on the width of potential drop. In Fig.2 (b), for $d=4.2x_0$, when $\theta$ varies from $0$ to $\pi$, $R_s$ for potential well increases initially, reaches the maximum value at $\theta=\pi/2$, and finally decrease to the initial value. While for $d=4.7x_0$, the variation of $R_s$ in terms of $\theta$ is completely opposite, which is similar to the case of potential barrier. This is in contrast with the reflection behavior from potential barriers (see Fig.2 (c) and Fig.3 (b)), which is immune against the width of potential barrier. The essential mechanism for such discrepancy will be demonstrated in later characteristic dynamics.

Through much simulation, we find that for both potential wells and potential barriers and in a certain regime of incident velocity, the asymptotic reflection probability always has the cosine form of the deflection angle between incident soliton and reflected soliton, i.e., $R_s$ $\sim $ $\cos 2\theta$. It is distinguished from the reflection of oblique incident waves, where the reflection probability is proportional to the normal velocity, e.g., $R_s(\theta) \sim \sin\theta$ for our system. This indicates that the interaction between reflected component and incident component of the matter-wave dark soliton may be an essential factor for the quantum reflection. We note that since the strong diffraction emerges for quantum reflection from potential drops (see Fig2.(a)\,(ii)), the deflection angle does not fulfill $2\theta$ exactly and corresponding $R_s{(\theta)}$ does not fit as perfectly as potential barriers (see Fig2.(a)\,(iii) and (c)).

\begin{figure}[htbp]
\begin{center}
\includegraphics[angle=0,width=0.45\textwidth]{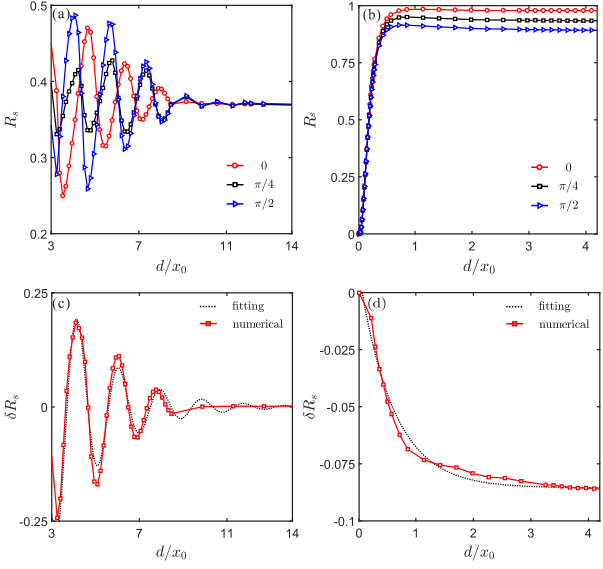}
\caption{(Color online) Top: The reflection probability, $R_s$, as the function of the widths of potential drop (a) and potential barrier (b).
Line-and-marker curves show the case of the condensate containing dark solitons with three different orientations for $\theta=0$ (red circles), $\theta=\pi/4$ (black squares) and $\theta=\pi/2$ (blue triangles).
Bottom: The variation of the reflection probability, $\delta R_s=R_s(\theta=\pi/2)-R_s(\theta=0)$, as a function of the widths of potential drop (c) and potential barrier (d). The solid curve with squares in (c) is from numerical calculation for $h=-180E_0$ and $\Delta_x=7.47x_0$, and dotted curve is the fitting from an oscillating exponential function, $\delta R_s = -0.74\exp{(-0.32d)}\sin{(3.42d-9.38)}$. Similarly, the solid curve with squares in (d) is from numerical results of the width of the barrier for $h = 40 E_0$, $\Delta_x = 7.05x_0$, and the dotted line is from the ansatz for potential barrier with $\delta R_s = 0.09\exp(-1.67d)-0.09$.
}
\label{fig:pz4}
\end{center}
\end{figure}

Fig.3 shows the influence of the width of rapidly varying potentials in the quantum reflection probability of the matter wave including a dark soliton. We find that when the width of the potential drop is less than the length of the soliton (or the condensate), i.e., $d<11.34x_0$, the reflection probabilities, $R_s$, for different orientation angle are sensitive to $d$ and displays oscillating exponential decay. While for $d>11.34x_0$, $R_s$ are immune against $d$ and approach identically a nonzero constant. For a potential barrier, since the kinetic energy of the incident wave is less than the barrier height, i.e., $ \frac{1}{2} m \omega^2(\Delta x)^2 < h$, there exists an exponential decay wave, $e^{-\alpha x}$ in the region of the potential barrier. Here $\alpha$ is approximately equal to $\sqrt{2m(V-E)/\hbar^2}=5.5/x_0 $ for Fig.3 (b) and (d), it indicates the wave decays nearly zero when $x=x_0$, i.e., $e^{-\alpha x_0}=0.0041$, which matches Fig.3 (b) well where $R_s$ tends to a stable value when $d>x_0$. Therefore, we argue that when d is less than the decay length of the matter wave in the barrier, the orientation of the dark soliton has trivial effect on $R_s$ while becomes significant when $d$ is larger than the decay length. Moreover, Fig.3 (c) and (d) show that the variation of reflection probabilities $\delta R_s=R_s(\theta=\pi/2) - R_s(\theta=0)$ for two system display distinct behaviors: $\Delta R_s$ shows an oscillating exponential decay into \emph{zero} for a potential drop while displays an exponential decay into a \emph{nonzero} constant for a potential barrier. It can be also seen that the quantum reflection of matter-wave dark soliton from the potential drop is very sensitive to the small region of potential length and $\delta R_s$ can varies up to $0.25$ ($1$ means a total reflection), while the reflection from the potential barrier is insensitive to small $d$ and $\delta R_s$ decreases monotonically to a constant with increasing $d$. This indicates that there exist two mechanism for the quantum reflection of the condensate containing a dark soliton \cite{PhysRevA.81.043610}: for potential drops, the length of matter-wave dark soliton determines the sensitive width of potential drops; for potential barriers, the decay length of the matter wave in the region of the barrier qualifies the sensitive width of the barrier.

\begin{figure}[htbp]
\begin{center}
\includegraphics[angle=0,width=0.49\textwidth]{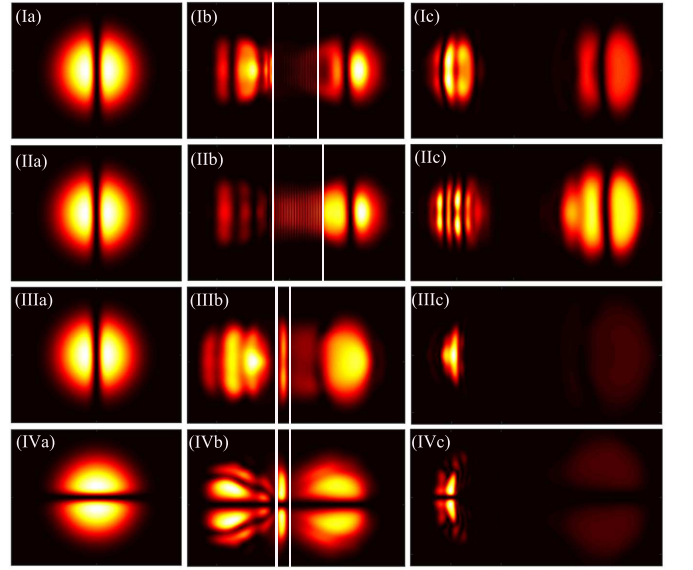}
\caption{Top two rows: The temporal density distribution of the BEC containing a $\theta=\pi/2$ dark soliton and reflected from potential drop of different widths. The depth of the potential well is $h=-180E_0$, and the widths of the potential well in (Ia)-(Ic) and (IIa)-(IIc) are $d=4.2x_0$ and $d=4.7x_0$, respectively.
Bottom two rows: The temporal density distribution of the BEC containing a dark soliton with orientation angle $\theta=\pi/2$ and $\theta=0$ interacting with the potential barrier, respectively. Here, its width is $d=0.28x_0$ and its height is $h=100E_0$. The regime between two solid white lines in (Ib) and (IIb) as well as (IIIb) and (IVb) indicate the case of potential well and the barrier, respectively. The points of time are $t= 0.00t_0$ ((Ia), (IIa), (IIIa), (IVa)), $t = 1.12t_0$ ((Ib), (IIb)), $t=1.84t_0$ ((Ic), (IIc)), $t=2.40t_0$ ((IIIb),(IVb)) and $t=3.68t_0$ ((IIIc), (IVc)).
}
\label{fig:pz4}
\end{center}
\end{figure}

Conventional quantum physics shows that the incident wave with the energy $E$ and velocity $\sqrt{2mE/\hbar}$ leads approximatively to the transmitted wave with velocity $\sqrt{2m(E-V)/\hbar}$. Here $V>0$ for potential barriers and $V<0$ for potential drops. For potential drops, $V < 0$ and $E - V \gg E$ in our simulation, indicating that the velocity of the transmitted wave is much larger than the velocity of the incident wave as well as the reflected wave. Due to the interatomic interaction and the absence of confinement potential in the area of drop, the transmitted wave can expand significantly in a short time. As a result, an interference pattern with very small fringe spacings occurs in a relatively large condensate wave packet. This is in contrast with the system of potential barriers, where the small condensate wave packet contains an interference pattern with large fringe spacings. Furthermore, the interatomic interaction causes the change of the reflected component of the condensate in shape, which is very different from the incident component. Also the transmitted component distinguishes from the reflected component, so the matter wave reflection is not a strict specular reflection.

To explore the discrepancy of the reflection behavior in the potential drop and potential barrier, we study the time evolution of the condensate density distribution as shown in Fig.4. For $d=4.2x_0$, the transmitted wave and the reflected wave in the region of potential drop (see the bracketed area by two white lines in Fig.4(Ib)) produce an interference pattern with small fringe spacings. Due to the proper width of potential drop and interatomic interaction, the valley of the matter-wave packet, i.e., a low-density area, emerges in the drop region and increases the effective depth of the potential drop, thus enhancing the quantum reflection. By contrast, for $d=4.7x_0$, the interference of the matter wave causes the peak of the wave package, i.e., a high-density of area, to occur in the region of potential drop (see Fig.4 (IIb)), which reduce the effective depth of the potential drop and thereby suppress the quantum reflection. Therefore, it demonstrate that the interference pattern as well as interatomic interaction are critical factors for the quantum reflection of the condensate containing a dark soliton from the potential drop. Moreover, for the reflection of the BEC from potential barrier, the transmitted velocity in the region of barrier (see the bracketed areas in Fig.4(IIIb) and (IVb)) are much smaller than the incident velocity during the transmission process, so there do not exist wave knots and complex interference pattern. As a result, the reflection behavior is immnune against the small barrier width. But for different orientation angle of the dark soliton, the quantum reflection for both systems is still sensitive. The bottom two rows show the reflection process of the condensate with different orientation from the same potential barrier. When the soliton orientation is parallel to the barrier surface (Fig.4(IIIa)) and during the reflection process, the soliton structure tends to prevent the return of the reflected wave by barriers due to its self-reinforcement and thus reduces the reflection rate (Fig.4(IIIb)). Note that there exist two reflected waves from the potential barrier. While for $\theta=0$ the influence of the dark soliton on the reflected waves are trivial and closer to the case of no soliton (see Fig.2 in \cite{Qiao:2018a}) and thereby the reflection rate will increase. Therefore, when a BEC containing a dark soliton is reflected from a potential barrier, the reflection probability for $\theta=\pi/2$ is smaller than the one for $\theta=0$ (see Fig.2(c)).
From Fig.4 ((Ic), (IIc)), one can also see that after the condensate contacts with the potential well, the transmitted component of the matter wave contains a dark-soliton structure with the life time longer than the reflected component, indicating the stability of the dark soliton passing through a potential drop as topological excitation.

\begin{figure}[htbp]
\begin{center}
\includegraphics[angle=0,width=0.45\textwidth]{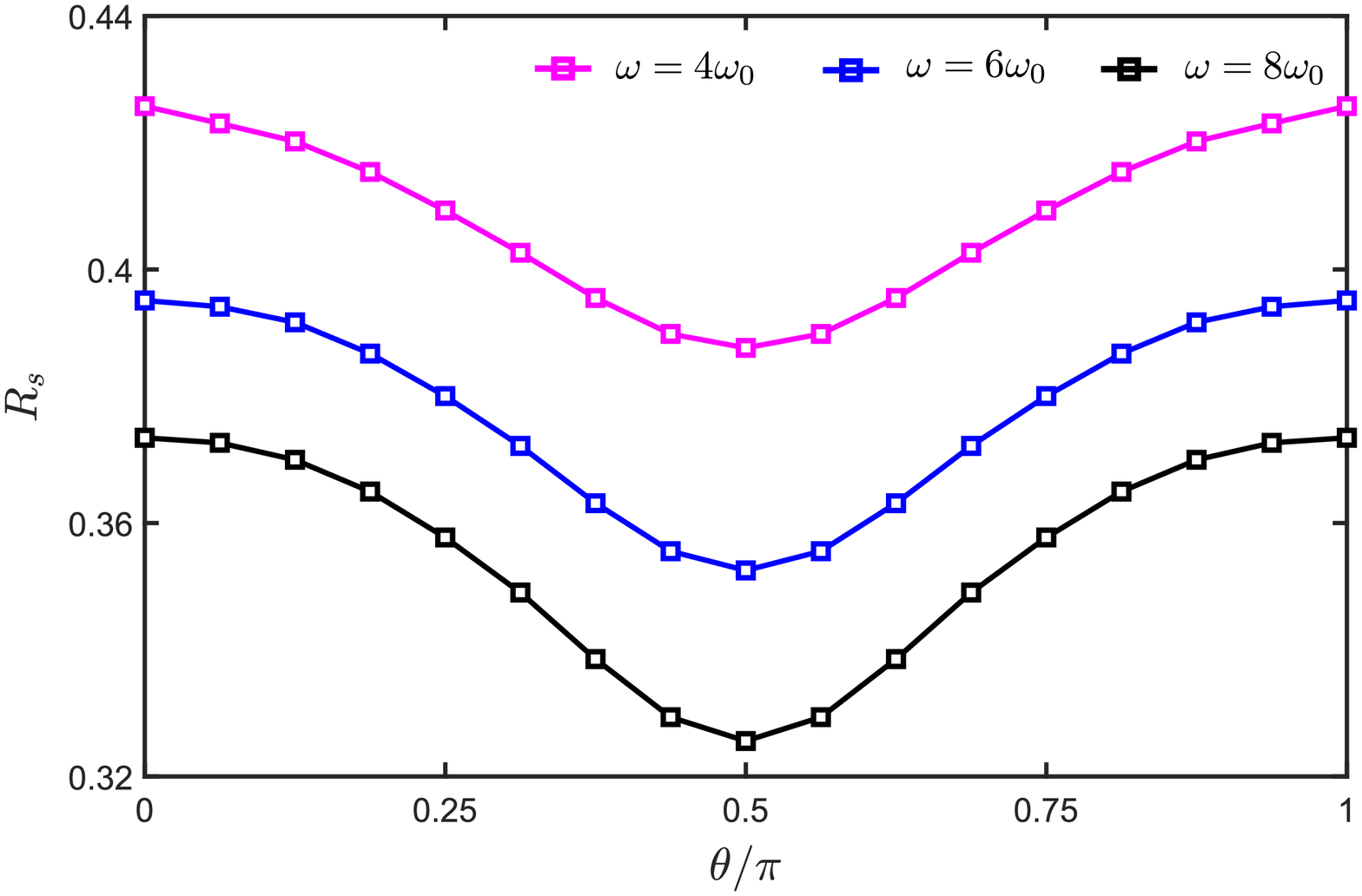}
\caption{(Color online) The reflection probability of the BEC  dominated by expansion motion as a function of the orientation of the soliton. Different line-and-maker plots indicate the condensate in different initial traps: $\omega=4\omega_0$ (squared line); $\omega = 6\omega_0$ (circular line); $\omega=8\omega_0$ (diamond line)}
\label{fig:pz4}
\end{center}
\end{figure}

Next we explore the effect of the expansion motion of the condensate containing a dark soliton with different orientation angle on its quantum reflection. To obtain the expansion-dominated BEC, we create BEC under different initial traps: $\omega=4\omega_0$, $\omega=6\omega_0$ and $\omega=8\omega_0$. At time $t=0$, we suddenly displace the harmonic trap by a distance $2.83 x_0$  and change its trap frequency to $\omega_0$. Thus the condensate expands rapidly when it is accelerated toward the potential well. Fig.3 shows the reflection probability of the expansion-dominated BEC, $R_s$, with respect to different orientation angle $\theta$. It can seen clearly that the reflection rate is still sensitive to the orientation of the dark soliton; the variation of $R_s$ is up to $0.05$. The behavior of expansion motion on $R_s$ is opposite to the one of c.m. motion on $R_s$. In the expansion-dominated process, from $\theta=0$ to $\pi/2$, $R_s$ here decreases gradually and approaches the minimum value at $\theta=\pi/2$. After $\theta$ varies from $\pi/2$ to $\pi$, $R_s$ increase gradually. Moreover, the larger expanding speed, the smaller $R_s$for the same $\theta$.

\section{Experimental parameters and Summary}

With regard to the matter-wave dark soliton with adjustable orientation angles, it may be prior to produce it by combining the condensate in a double potential well with the phase imprinting \cite{PhysRevLett.101.120406, Denschlag:1999a, Meyrath:2005a}. The sensitive widths of potential wells and barriers for the dark-soliton condensate are approximately the length of the soliton and the decay length of the matter wave in the region of potential barrier, respectively. According to our system, where a quasi-2D condensate is produced under the parameters $\omega=2\pi \times 10 Hz$, $\omega_z=2\pi \times 100 Hz$, $N=1.8 \times 10^4$, the length of the dark soliton is $11.34 x_0 \approx 75\mu m$ and the decay length is approximately the order of $x_0\approx 6.6\mu m$. Curret cold-atom experiments can achieve boxlike potential with length varing from several micrometers to hundreds micrometers \cite{PhysRevLett.110.200406, PhysRevA.71.041604}, so the potential drops can be realized in current experiments. Also we can increase the size of the condensate via the trap potentials and interatomic interaction to fulfill the experimental requirment.

In summary, we have investigated some crucial factors that influence the reflection probability of a dark-soliton condensate. In particular, the widths of the potential well and barrier have a great influence on the change of reflection. They are determined, respectively, by the length of the dark soliton and the decay length of the matter wave in the region of the potential barrier. Furthermore, we find that for a rapidly changing potential, the quantum reflection probability is represented as the cosine of the deflection angle between the incident soliton and the reflected soliton within a certain range of incident velocity, namely $R(\theta)$ $\sim $ $\cos 2\theta$. The sensitivity of reflection probability on the orientation of soliton may permit experiments to detect the existence of a dark soliton and probe the possible structure of soliton. Since the BEC with a dark-soliton profile is much susceptible to potential drops than potential barrier for a certain width up to the order of  the width of the condensate, this suggests that a potential drop may be a desirable candidate for new devices used to detect the nonlinear excitation. Our results may provide a general view about the micro-mechanism of atoms on solid surface potential, for example, semiconductor surfaces, graphene surfaces or nanostructures.

\acknowledgments
We thank Y.-J.Lin for the inspiriting discussion about experimental setup for our system. This work is supported by the NSFC under grants Nos. 11775178 and 11775177, the Major Basic Research Program of Natural Science of shaanxi Province under grant Nos. 2017KCT-12 and 2017ZDJC-32, and the Open Research Fund of Shaanxi key Laboratory for Theoretical Physics Frontiers under grant No. SXKLTPF - K20190602.

%
%

\bibliography{Refs}

\end{document}